\documentclass[french]{hermes-journal}

\usepackage[utf8]{inputenc}
\usepackage{graphicx}
\usepackage{scalerel}
\usepackage{tikz}
\usepackage{footnote}
\usepackage{adjustbox}
\usepackage[]{algorithm2e}
\usetikzlibrary{svg.path}

\definecolor{orcidlogocol}{HTML}{A6CE39}
\tikzset{
  orcidlogo/.pic={
    \fill[orcidlogocol] svg{M256,128c0,70.7-57.3,128-128,128C57.3,256,0,198.7,0,128C0,57.3,57.3,0,128,0C198.7,0,256,57.3,256,128z};
    \fill[white] svg{M86.3,186.2H70.9V79.1h15.4v48.4V186.2z}
                 svg{M108.9,79.1h41.6c39.6,0,57,28.3,57,53.6c0,27.5-21.5,53.6-56.8,53.6h-41.8V79.1z M124.3,172.4h24.5c34.9,0,42.9-26.5,42.9-39.7c0-21.5-13.7-39.7-43.7-39.7h-23.7V172.4z}
                 svg{M88.7,56.8c0,5.5-4.5,10.1-10.1,10.1c-5.6,0-10.1-4.6-10.1-10.1c0-5.6,4.5-10.1,10.1-10.1C84.2,46.7,88.7,51.3,88.7,56.8z};
  }
}

\newcommand\orcidicon[1]{\href{https://orcid.org/#1}{\mbox{\scalerel*{
\begin{tikzpicture}[yscale=-1,transform shape]
\pic{orcidlogo};
\end{tikzpicture}
}{|}}}}

\firstpagenumber{1}

\title[Quelle blockchain choisir ?]{Quelle Blockchain choisir ? Un outil d’aide à la décision pour guider le choix de technologie Blockchain}

\author{Nicolas}{Six}
\author{Nicolas}{Herbaut}
\author{Camille}{Salinesi}

\address{Centre de Recherche en Informatique – EA 1445 \\
Université Paris 1 Panthéon-Sorbonne}
        {nicolas.six,nicolas.herbaut,camille.salinesi@univ-paris1.fr}

\abstract{
        Companies trying to build new solutions using blockchain are confronted 
        with a plethora of available concurrent technologies that have many control knobs which require fine-tuning by experts.
        Exiting studies that build decision models for blockchain adoption or selection lack an automated way to use non-functional requirements to provide recommendations.
        In this paper, we build a knowledge base for blockchain solutions by analyzing 
        whitepapers and studies, but also our benchmark results performed 
        in a controlled environment. 
        Then, we implement a Multi-Criterion Decision Analysis method to determine 
        the most suitable blockchain solution from companies provided requirements and preferences. 
        Finally, we illustrate our approach by running the decision process on a realistic 
        supply-chain use case.
        This paper provides a rationale for blockchain deployment 
        choices. While still limited in scope, we plan to include more blockchain 
        alternative and more flexible requirements inputs in future work.
}

\resume{
        Les entreprises qui souhaitent déployer des solutions basées sur la blockchain sont confrontées à une pléthore de technologies concurrentes ayant chacun un grand nombre de paramètres propres devant être ajustés par un expert.
        Dans cet article, nous proposons un algorithme d'aide à la décision permettant de mieux prendre en compte les exigences haut niveau et préférences pour les recommandations.  
        Nous construisons une base de connaissance de solutions blockchain et exécutons un processus de décision multicritère automatisé donnant la solution la plus pertinente à partir des exigences et préférences. 
        Nous validons notre approche sur un cas d’étude de gestion de chaîne logistique. 
        Nous donnons des pistes pour une prise en compte plus flexible des exigences dans de futurs travaux.        
}

\keywords{
        Blockchain, 
        Requirements engineering, 
        Multi-criteria decision analysis
}
\motscles{
        Blockchain, 
        Ingénierie des exigences, 
        Aide à la décision multicritère
}

\begin{document}

\maketitle

\newpage

\section{Introduction}

La blockchain (ou chaîne de blocs) est un registre distribué maintenu à jour par un ensemble de nœuds. Les utilisateurs peuvent interagir avec les nœuds afin d'envoyer à la blockchain des transactions. 
Un registre blockchain prend la forme d'un ensemble de blocs contenant les transactions soumises par les utilisateurs, ainsi que des métadonnées sur lui-même ou le registre. Chaque bloc est relié au précédent par la valeur de hachage (en anglais, \textit{hash}) de celui-ci. Dans ce sens, comme la modification d'un bloc altérerait cette valeur ainsi que toutes les autres valeurs de hachage des blocs suivants, il est théoriquement impossible d'altérer le contenu d'un bloc. 
Les nouveaux blocs sont formés par un sous-ensemble de nœuds chargés de regrouper les transactions dans un bloc et de le valider en mettant en œuvre différents processus cryptographiques en fonction de la blockchain utilisée, afin de garantir sa validité lors de son ajout à la blockchain. 

La blockchain est apparue lors de la création de Bitcoin \cite{nakamoto2008bitcoin}, afin de permettre aux utilisateurs d'échanger la cryptomonnaie de même nom. Par la suite, de nombreuses blockchains ont pu voir le jour. Ethereum, la plus connue, est plébiscitée \cite{wood2014ethereum} pour sa capacité à déployer et à interagir avec des contrats intelligents (en anglais, smart contracts), logés dans la blockchain \cite{Szabo1997}. 
Les contrats intelligents pour la blockchain permettent non seulement d'exécuter des fonctions directement au sein de celle-ci, mais aussi de stocker des états. 
Ils bénéficient donc directement des propriétés particulières de la blockchain, qui sont intégrité, décentralisation, non-répudiation des transactions et transparence. 
Cela donne à la blockchain un statut de tiers de confiance artificiel, où il est possible de faire confiance au code et à la puissance du réseau contrairement aux tiers de confiance conventionnels qui assurent la validité des transactions au travers de leur statut, tels que les banques ou les gouvernements. 

Ces caractéristiques ont attiré l'attention des industriels et universitaires, qui voient en la blockchain un moyen de révolutionner la manière d'échanger de la valeur entre individus ainsi que de garantir la véracité et l'intégrité des données stockées dans celle-ci. 
En effet, la blockchain serait "un support numérique natif pour la valeur, par lequel nous pourrions gérer, stocker et échanger de multiples biens [...] de pair à pair et de manière sécurisée" \cite{dontapscott2016}. 
De ce fait, on trouve dans la littérature de nombreux cas d'utilisations pertinents de la blockchain dans différents secteurs d'activités, tels que la gestion de chaîne logistique \cite{abeyratne2016blockchain}, la finance \cite{hyvarinen2017blockchain}, le contrôle du réseau \cite{herbaut2017model}, l'identité décentralisée numérique \cite{takemiya2018sora} ou encore la santé \cite{ekblaw2016case}.

Cependant, malgré son potentiel, la blockchain fait face à de nombreux freins à l'adoption.
D'après une étude réalisée par l'entreprise PwC en 2018\footnote{https://www.pwc.com/gx/en/issues/blockchain/blockchain-in-business.html}, les entreprises se heurtent à des problèmes tels qu'une régulation incertaine sur le sujet, un manque de confiance envers les autres acteurs lors de leur participation à un projet utilisant la blockchain, ainsi qu'à la difficile gestion de la propriété intellectuelle des données et biens qui y sont enregistrés. Ces problèmes sont résolus petit à petit grâce à la collaboration des acteurs de l'écosystème blockchain et des instances juridiques et gouvernementales compétentes. 
Néanmoins, les entreprises sont encore confrontées à un frein technologique, et ce pour plusieurs raisons. 
Elles peuvent rencontrer des difficultés à recruter des collaborateurs spécialisés en blockchain, la technologie étant encore jeune. 
Elles peuvent aussi avoir du mal à intégrer la blockchain à leurs systèmes d'information et processus métiers existants, car il n'existe pas encore de bonnes pratiques identifiées et éprouvées en entreprise par les architectes logiciels.
Afin de pallier ce problème, des études ont été menées afin d'assister l'intégration de la blockchain dans des architectures logicielles.
Dans ce sens, une étude propose une collection de modèles architecturaux contenant de la blockchain, ainsi que les différents cas où ces modèles sont applicables \cite{Xu2018}. 

Mais le frein principal se situe sur la conception de la solution blockchain ainsi que son implémentation. 
À ce stade, les développeurs peuvent se poser plusieurs questions.
Quelle blockchain utiliser dans un contexte donné, sachant qu'il existe de nombreuses technologies concurrentes avec, pour chacune, des propriétés et caractéristiques qui leur sont propres ? 
Peut-être est-ce finalement plus raisonnable d'utiliser une solution "éprouvée" au lieu d'une blockchain (base de données, microservices ...) ? 
Enfin, comment configurer les différents paramètres de la blockchain, qui ont un impact important sur la satisfaction des exigences (performances, résilience, sécurité ...) tels que l'algorithme de consensus ou l'intervalle inter-blocs, nécessitant souvent l'intervention d'experts dans le domaine pour aboutir à un résultat satisfaisant les exigences ?

De nombreuses études ont été menées pour répondre aux deux premières questions et ainsi facilitent le choix de la solution blockchain, notamment par le biais de modèles de décision à travers diverses questions \cite{Wust2018, Koens2018}.
Une autre étude \cite{Belotti2019} présente un \textit{vadémécum} contenant toutes les informations nécessaires à la compréhension de la blockchain d'un point de vue technique, ainsi qu'un modèle de décision pour la blockchain appliqué à plusieurs scénarios d'exemple. 
Les systèmes de recommandation proposés sont souvent constitués d'une série de questions abstraites, permettant de répondre à des problématiques telles que "ai-je besoin d'une blockchain ?" ou "quel type de blockchain adopter ?", mais pas de fournir des recommandations précises, ou rentrent plus dans le détail en considérant les choix entre de nombreux paramètres et propriétés blockchain. Les utilisateurs souhaitant obtenir une recommandation plus précise doivent donc se tourner vers ces derniers.
Ce type d'étude est pertinent pour des personnes ayant de bonnes connaissances dans le domaine de la blockchain, mais il sera difficile pour des personnes non initiées de répondre de manière précise aux questions du modèle de décision. 
De plus, beaucoup de ces études se concentrent uniquement sur les exigences blockchain, alors que les utilisateurs ont des exigences liées à la qualité logicielle (performance, sécurité, fiabilité ...). Les liens qui relient les attributs blockchain aux qualités logicielles telles que définies en ingénierie, sont souvent peu explicites et il est difficile de quantifier l'impact d'un paramètre blockchain sur les qualités logicielles de la solution finale. 
Enfin, lorsque le nombre d'attributs techniques considérés devient important, il est impossible de réaliser un choix les prenant tous en compte, la complexité de calcul lors de l'utilisation manuelle du modèle étant trop élevée. 

Pour pallier ces limitations, nous introduisons dans cet article un processus de décision automatisé qui détermine l'alternative la plus intéressante pour un cas d'étude donné.
Dans celui-ci, les préférences ou exigences des utilisateurs quant à la qualité logicielle de la solution à créer seront utilisées en entrée.
Celles-ci seront comparées aux différentes caractéristiques des alternatives considérées par une méthode d'aide à la décision multicritère.
Ces caractéristiques seront contenues sous la forme d'une base de connaissance et définies en utilisant la littérature existante (expériences, revues de littérature ...), les livres blancs des blockchains considérées ainsi que nos propres résultats d'expériences. 
Nous présentons également une application de notre processus de décision à un cas d'utilisation pertinent dans le domaine de la gestion de chaîne logistique. 
Cette partie sera l'occasion pour nous de valider les résultats du processus de décision, par le biais d'expériences manuelles confirmant les décisions prises par le processus. 

La section~\ref{inputs} de cette étude est consacrée au processus de décision, la section~\ref{casestudy} sur l'application du processus au cas d'étude sur la gestion de chaîne logistique. Nous présentons les travaux connexes à notre étude dans la section \ref{relatedwork}, puis nous enchaînons sur une discussion quant à nos résultats et notre approche dans la section~\ref{discussion}. Enfin, nous concluons notre étude et introduisons nos travaux futurs dans la section~\ref{conclusion}.

\section{Construction du processus de décision \label{inputs}}

Dans cette section, nous allons présenter les entrées ainsi que le fonctionnement du processus de décision permettant d'aider l'utilisateur à choisir le modèle de blockchain le plus approprié.
\subsection{Entrées}

La précision d'un algorithme d'aide à la décision multicritère dépend majoritairement des données saisies en entrée.
Dans cette sous-section, nous présentons notre approche pour construire une base de connaissance fiable et adaptée, ainsi que notre méthode pour éliciter les poids qui seront appliqués à chacun des critères pour l'exécution du processus de décision.

\subsubsection{Alternatives et attributs}

Pour alimenter notre processus d'aide à la décision, nous avons construit une première version de base de connaissance contenant un ensemble d'alternatives de blockchains $a_{m}$ et de leurs attributs respectifs $c_{n}$ (Tableau~\ref{knowledge}).
Nous avons choisi ce panel spécifique de blockchains, car (hors Bitcoin) elles sont considérées comme les blockchains les plus utilisées par les fournisseurs de service blockchain en entreprise\footnote{https://www.hfsresearch.com/pointsofview/whos-winning-the-battle-of-enterprise-blockchain-platforms}. Cependant, nous avons quand même choisi d'inclure la blockchain Bitcoin dans notre base de connaissance, car elle la plus connue du grand public, mais aussi la plus ancienne.

L'objectif de notre travail étant d'aider les entreprises à prendre des décisions sur la blockchain à utiliser sans avoir d'expertise particulière quant à la configuration de celle-ci, nous avons choisi un ensemble de critères qui peuvent être placés sous les différents macro-attributs proposés par la norme ISO 25010\footnote{https://iso25000.com/index.php/en/iso-25000-standards/iso-25010}, un standard définissant les différents macro-attributs à considérer afin de garantir la qualité d'un système ou d'un logiciel lors de son implémentation. 
Nous avons choisi les attributs qui nous semblent pertinents dans les considérations à avoir lors du choix d'une blockchain, mais aussi pour la possibilité à les retranscrire sous format numérique.
Par conséquent, nos critères ne sont pas spécifiques à la technologie blockchain, mais relatifs à la qualité système. C'est notre processus de décision qui aura la charge de transcrire ces attributs de qualité système en attributs blockchain (tels que l'intervalle inter-blocs, l'algorithme de consensus, ou la taille des blocs).
La figure \ref{attr-map} présente un diagramme indiquant les relations entre les attributs de qualité logicielle (critères choisis pour notre processus de décision) et les attributs spécifiques à la blockchain.
\begin{figure}[attr-map]{Attributs choisis (milieu) reliés aux qualités système (à gauche) et blockchain (à droite).}
    \centering
    \includegraphics[scale=0.35]{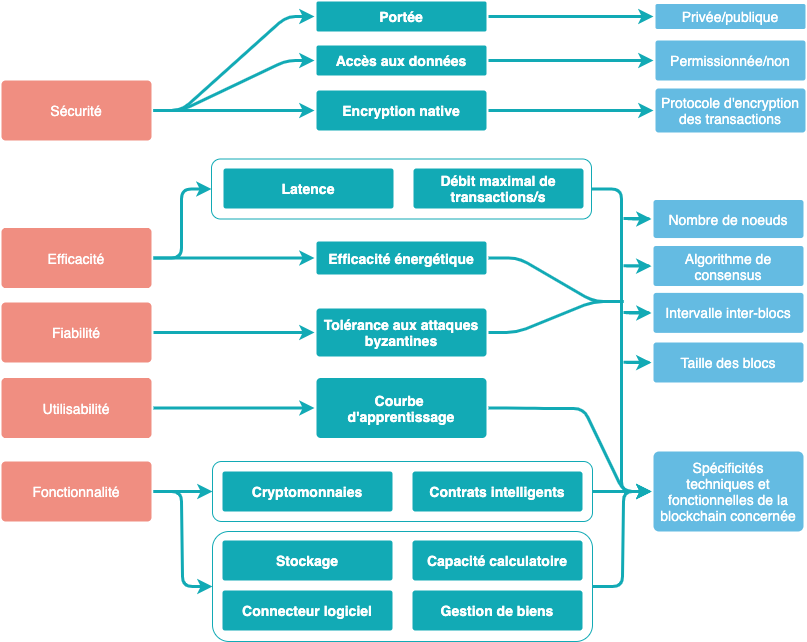}
\end{figure}
Les valeurs saisies pour chacun des attributs d'alternatives de notre base de connaissance proviennent de différentes sources : études (comme celle de \cite{Belotti2019}), livres blancs (e.g. \cite{brown2016corda}, \cite{nakamoto2008bitcoin}, \cite{wood2014ethereum}), documentations techniques et littérature scientifique (e.g. \cite{androulaki2018hyperledger}).

Certaines de ces valeurs sont approximatives (marquées par le symbole $\mp$), car soumises à variation, de la topologie et la configuration du réseau blockchain ainsi que des caractéristiques techniques des nœuds le composant (CPU, mémoire vive ...).
Leur valeur est donc construite à partir d'attributs connus, comme l'algorithme de consensus supporté (un algorithme tolérant les fautes byzantines comme l'algorithme PoW de Bitcoin aura un débit de transaction plus faible qu'un algorithme tolérant aux pannes, comme Raft utilisé par Hyperledger Fabric). 
Néanmoins, ces valeurs peuvent être fixées lorsque les paramètres blockchain sont connus.
Notre processus de décision devant prendre en compte des actifs déjà présents dans l'entreprise (comme  l'infrastructure technique ou les modèles de processus métiers), nous comptons sur la réalisation de tests de performance afin de pouvoir donner une valeur fixe aux attributs variables en fonction du contexte donné. 
Cette base de connaissance sera également variable dans le temps. Les valeurs des attributs des différentes blockchains choisies seront modifiées si nécessaire (mise à jour d'un des éléments d'une blockchain). Ces variations pouvant avoir un impact sur le choix de la meilleure alternative par notre processus de décision, il sera nécessaire d'évaluer l'ancienneté de la base de connaissance afin de déterminer si la recommandation est pertinente à un instant donné.
\begin{table}[knowledge]{Alternatives et attributs retenus.}
\centering
\begin{adjustbox}{width=1\textwidth}
\begin{tabular}{|r|l|l|l|l|l|}
\hline
\textbf{Attributs/Alternatives} & \textbf{Bitcoin} & \textbf{Ethereum} & \textbf{Ethereum} & \textbf{Hyperledger Fabric} & \textbf{Corda} \\
\hline
Algorithme de consensus & PoW\footnote{Proof-of-work (PoW), preuve de travail} & PoW & PoA \footnote{Proof-of-Authority (PoA), preuve d'autorité} & Raft & PBFT\footnote{Practical Byzantine Fault Tolerance} \\
\hline
\hline
Ouvert publiquement & Oui & Oui & Non & Non & Non \\
Permissions & Non & Non & Non & Oui & Oui \\
Encryption native & Non & Non & Non & Oui & Oui \\
\hline
Débit (tx/s) & 3,8 & 15 & $\mp100$ & $\mp1000$ & $\mp1000$ \\
Latence (s)& 3600 & 180 & $\mp10$ & \textless{}1 & \textless{}1 \\
Efficient en énergie & Non & Non & Oui & Oui & Oui \\
\hline
Tolérant aux fautes byzantines & 50,00\% & 50,00\% & 33,30\% & 0,00 \% & 33,30\% \\
\hline
Contrats intelligents & Non & Oui & Oui & Oui & Oui \\
Cryptomonnaies & Oui & Oui & Oui & Non & Non \\
Element de stockage & Basique & Avancé & Avancé & Avancé & Avancé \\
Elément de calcul & Non & Avancé & Avancé & Avancé & Avancé \\
Elément gestionnaire de biens & Basique & Avancé & Avancé & Avancé & Avancé \\
Connecteur logiciel & Non & Avancé & Avancé & Avancé & Avancé \\
\hline
Courbe d'apprentissage & Faible & Moyenne & Moyenne & Très élevé & Très élevé \\
\hline
\end{tabular}
\end{adjustbox}
\end{table}

\subsubsection{Poids et conditions définis par l'utilisateur}

Afin d'obtenir une préconisation de blockchain conforme aux attentes de l'utilisateur, le processus de décision automatisé doit prendre en compte les exigences et préférences de celui-ci. 
Lorsque l'utilisateur est invité à saisir ses choix, il peut marquer un critère comme \textit{Requis} ou \textit{Indésirable}. Lors de la prise de décision, une alternative dont l'attribut ne respecterait pas l'une de ces deux exigences serait automatiquement disqualifiée des alternatives possibles, indépendamment de son score obtenu par l'exécution de l'algorithme d'aide à la décision multicritère.

L'utilisateur peut également indiquer ses préférences quant aux attributs, par le biais de variables littérales formant une échelle de Likert \cite{allen2007likert} (Tableau \ref{likert}). Le choix d'une de ces variables permet d'obtenir une valeur de préférence $p_{n} \in \mathbb{n}$ pour chacun des critères $c_{n}$. Afin d'obtenir les poids de chacun des critères $\omega_{n}$ de telle façon à ce que la somme de ces poids soit égale à 1, il nous faut diviser chacune des préférences $p_{n}$ pour un critère par la somme des préférences.

\begin{table}[likert]{Échelle de Likert associant labels et valeurs de préférence.}
\centering
\begin{tabular}{|l|c|}
  \hline
  \textbf{Variable linguistique} & \textbf{Valeur de préférence $p_{n}$} \\
  \hline
  Extrêmement désirable & 4 \\
  Tout à fait désirable & 3 \\
  Désirable & 2 \\
  Faiblement désirable& 1 \\
  Indifférent & 0 \\
  \hline
\end{tabular}
\end{table}

\subsection{Logique interne}

Tout d'abord, notre processus de décision effectue un premier filtrage des alternatives en fonction des exigences de l'utilisateur. Si un critère marqué comme \textit{Requis} ou \textit{Indésirable} n'est pas respecté par l'une des alternatives, elle est automatiquement éliminée, peu importe le score qu'elle aurait pu obtenir à l'aide de l'algorithme de décision qui suit.
Pour un critère \textit{Requis} qui n'est pas un booléen, l'utilisateur  spécifie une valeur extremum.
Par exemple, si un certain nombre de transactions par seconde est requis, les alternatives qui n'atteignent pas la valeur seuil seront disqualifiées. 

Le processus de décision automatisé sur les alternatives restantes repose sur l'utilisation d'un algorithme d'aide à la décision multicritère appelé TOPSIS (Technique for Order of Preference by Similarity to Ideal Solution) \cite{lai1994topsis}. 
L'algorithme TOPSIS est basé sur le fait que l'alternative $a_{m}$ la plus pertinente pour un ensemble de choix donné doit être la plus proche possible de la solution idéale positive $A^+$ et la plus éloignée de la solution idéale négative $A^-$.

Le choix de cet algorithme a été guidé grâce à une étude présentant un état de l'art des études portant sur le choix d'une méthode d'aide à la décision multicritère \cite{kornyshova2007mcdm}. Celle-ci propose un cadre de décision incluant différentes propriétés sur lesquelles porter notre attention lors du choix d'une méthode d'aide à la décision multicritère. Nous avons jugé que la méthode TOPSIS était adaptée à notre processus de décision, notamment car elle supporte l'analyse multicritère d'attributs nombreux et variés (ce qui est le cas lors de la comparaison de deux blockchains) tout en étant simples d'implémentation et précise dans la décision. Elle permet également de prendre en compte des poids définis par un utilisateur, ce qui est requis étant donné le mode opératoire de notre processus de décision.
Plusieurs étapes sont nécessaires à l'exécution de la méthode TOPSIS, détaillées dans les sous-parties suivantes.

\textit{Construction de la matrice} - Soit $m$ alternatives $a$ et $n$ attributs $c$ pour chacune d'entre elles. Le regroupement de ces alternatives donne une matrice $X = \{x_{ij}\}$ pour $\{i \in \mathbb{N}\ |\ 1 \le i \le m \}$ et $\{j \in \mathbb{N}\ |\ 1 \le j \le n \}$.

\textit{Normalisation de la matrice et application des poids} - Normaliser les critères ayant des unités et échelles différentes entre eux est nécessaire afin de pouvoir les comparer entre eux. C'est également à cette étape que nous appliquons les poids provenant des préférences de l'utilisateur.

\begin{equation}
v_{ij} = r_{ij} * \omega_{j} = \frac{x_{ij}}{\sqrt{\sum_{i=1}^{m}{x^{2}_{ij}}}} * \omega_{j}
\end{equation}

\textit{Calcul des solutions idéales positive et négative puis mesure de l'écart avec chacune des alternatives} - En sélectionnant les meilleures et les pires performances de chacun des critères de la matrice de décision normalisée pondérée, on peut déterminer les solutions idéales positive $A^+$ et négative $A^-$ afin de mesurer l'écart de chacune des alternatives avec ces deux solutions que l'on notera $S^+$ et $S^-$. 

\begin{minipage}[t]{.45\textwidth}
\begin{equation}
    Pour A^+ = (v_{1}^+, ..., v_{j}^+),
\end{equation}
\begin{equation}
    S{i}^+ \triangleq  \sqrt{\sum_{j=1}^{m} (v_{ij} - v{j}^+)^2}
\end{equation}
\end{minipage}
\begin{minipage}[t]{.45\textwidth}
\begin{equation}
    Pour A^- = (v_{1}^-, ..., v_{j}^-),
\end{equation}
\begin{equation}
    S{i}^- \triangleq \sqrt{\sum_{j=1}^{m} (v_{ij} - v{j}^-)^2}
\end{equation}
\end{minipage}

\textit{Calcul de la distance relative avec la solution idéale} - Cette dernière étape permet de donner un score à chaque alternative, qui représente sa distance avec la solution idéale. 
L'ordonnancement de ces scores permet de définir la meilleure alternative possible par rapport aux alternatives données ainsi que des préférences de l'utilisateur. 

\begin{equation}
    C{i} = \frac{S{i}^-}{S{i}^+ + S{i}^-}
\end{equation}

\section{Application à un cas d'étude de gestion de chaîne logistique \label{casestudy}}

Afin de tester le processus de décision automatisé proposé, nous avons sélectionné une étude qui propose d'introduire un système blockchain à une chaîne logistique afin de permettre le partage de données entre les différents acteurs \cite{longo2019blockchain}. Dans cette partie, nous détaillerons le scénario proposé par l'étude citée, puis les différents attributs requis pour la blockchain à implémenter qui découlent de ce sujet afin d'exécuter notre processus de décision automatisé. Enfin, nos résultats sont validés en utilisant un outil permettant de tester ses performances implémenté dans ce but.

\subsection{Scénario "Big-Box"}


La chaîne logistique modélisée dans cette étude est constituée d'un réseau de détaillants de la chaîne de magasins Big-Box, ainsi que de trois grossistes qui alimentent leurs magasins. 
Les détaillants Big-Box étant regroupés dans une même organisation, l'étude considère qu'il y a un partage de données en temps réel, transparent et fiable entre les magasins.
Cependant, les détaillants sont tout de même en compétition, car ils opèrent dans une même région géographique et proposent tous les mêmes gammes de produits.
Les clients arrivent au magasin et sélectionnent des produits ainsi que leurs quantités respectives. 
Si le stock du magasin permet de satisfaire la demande, le produit concerné est réservé dans la quantité demandée; sinon, une réservation partielle est proposée; la demande non satisfaite est utilisée pour calculer les réapprovisionnements.
L'inventaire est fait avant l'ouverture des magasins; si une commande est nécessaire alors le détaillant peut choisir l'un des grossistes pour s'approvisionner, en prenant en compte le délai d'approvisionnement, la demande actuelle et la quantité instantanément disponible pour les produits souhaités. 
Si la quantité d'un produit possédée par un grossiste devait ne pas être suffisante pour tous les détaillants, alors celui-ci est partagé équitablement. 

Dans ce contexte, partager la demande globale des différents détaillants entre les grossistes pourrait permettre de prédire plus facilement le stock à constituer pour répondre aux demandes des détaillants. 
Cependant, les acteurs de ce système restent en concurrence et ne se font donc pas confiance mutuellement. 
Les auteurs de l'étude proposent donc la mise en place d'une blockchain permettant d'y enregistrer des données liées à la chaîne d'approvisionnement (notamment la demande du marché) sous la forme de valeur de hachage, ainsi que les différents tiers ayant accès à ces données (s'ils sont autorisés par la blockchain, ils peuvent faire directement une requête pour obtenir ces données auprès du tiers qui les a enregistrées). 
La sauvegarde de cette valeur permettant d'attester de la véracité des données transmises entre tiers, ils peuvent dorénavant se faire confiance entre eux. 

\subsection{Exigences du client Big-Box}

Pour pouvoir préconiser la blockchain à l'aide de notre processus de décision, il convient d'identifier les attributs de qualité ainsi que les exigences et préférences quant à ces attributs (Section \ref{inputs}). Cette sous-section aborde donc chacun des attributs de qualité système proposés précédemment et explique le choix de la valeur de chacun d'entre eux. 

\textit{Sécurité} - Les données stockées étant hachées, elles ne sont pas considérées comme sensibles, pas plus que l'identité des tiers qui est masquée par leur adresse. Il est donc possible d'utiliser une blockchain publique (ce qui est d'ailleurs le choix initial de l'étude), sans chiffrement des données. Les permissions étant par ailleurs gérées à l'échelle du contrat intelligent, il n'est pas nécessaire d'avoir une blockchain supportant la gestion de permissions. Par déduction, ces propriétés n'étant pas importantes dans ce contexte, elles sont toutes marquées comme étant \textit{Indifférent} dans notre tableau d'entrées.

\textit{Efficacité} - Le système blockchain n'a pas besoin de supporter un débit minimal de transactions par seconde (que l'on différencie du nombre de transactions par seconde supportable soumises en entrée) ainsi qu'une latence particulière. Néanmoins, une latence faible pouvant être profitable à l'expérience utilisateur, nous avons tout de même choisi de la fixer à \textit{Faiblement désirable}. 
Pour ce qui est de l'efficience énergétique, c'est une propriété particulièrement intéressante dans une optique de réduction de coûts. L'utilisation de blockchains publiques à algorithme de consensus lourds (tel que PoW) est très coûteux en énergie. Nous avons donc choisi la préférence \textit{Tout à fait désirable} pour cette propriété.  

\textit{Fiabilité} - Les acteurs ne se faisant pas confiance entre eux, il est indispensable d'avoir un pourcentage de tolérance aux fautes byzantines, ce qui indique que le système est capable de fonctionner correctement pour un certain nombre de nœuds pouvant avoir un comportement adverse. Nous avons choisi un pourcentage d'au moins 33,3\%, ce qui permet de garantir la bonne continuité du réseau blockchain pour un nombre de nœuds fautifs $f + 1 < \frac{n}{3}$, $n$ étant le nombre de nœuds totaux constituant le réseau. 

\textit{Fonctionnalités} - Pour répondre aux objectifs du sujet défini, la blockchain doit être capable de prendre la forme d'un élément de stockage pour contenir les données des détaillants ainsi que de supporter l'administration de celles-ci, de facto par le biais de contrats intelligents. Ces deux attributs sont donc définis respectivement comme \textit{Avancé} ainsi que \textit{Requis}. Les autres fonctionnalités n'étant pas nécessaires, elles sont marquées \textit{Indifférent}.

\textit{Utilisabilité} - Enfin, le dernier attribut choisi est la courbe d'apprentissage : dans un contexte où la blockchain doit permettre d'économiser des coûts associés à la chaîne d'approvisionnement ainsi que de supporter une application de faible complexité, utiliser une technologie dont il est facile d'en apprendre les mécaniques peut être un atout. Nous avons choisi de le marquer \textit{Désirable}.

\subsubsection{Compilation des valeurs choisies}

La compilation des valeurs de ces qualités système aboutit au tableau~\ref{scexigences}, nous permettant d'exécuter notre processus de décision automatisé dans ce contexte.

\begin{table}[scexigences]{Exigences du système blockchain souhaité pour le cas d'étude.}
\centering
\begin{adjustbox}{width=\textwidth}
\begin{tabular}{|r|l|l|l|}
\hline
\textbf{Attributs} & \textbf{Exigences} & \textbf{Valeur exigée} & \textbf{Préférences}  \\
\hline
Ouvert publiquement & Aucune &  & Indifférent \\
Permissions & Aucune &  & Indifférent \\
Encryption native des données & Aucune &  & Indifférent \\
\hline
Débit (tx/s) & Aucune &  & Indifférent \\
Latence (s) & Aucune &  & Faiblement désirable \\
Efficient en énergie & Aucune & & Tout à fait désirable \\
\hline
Tolérant aux fautes byzantines & Requis & $\ge$33,33 \% & Désirable \\
\hline
Contrats intelligents & Requis & Oui & Indifférent \\
Cryptomonnaies & Aucune & & Indifférent \\
Élément de stockage & Requis & Avancé & Indifférent \\
Élément de calcul & Aucune & & Indifférent \\
Élément gestionnaire de biens & Aucune & & Indifférent \\
Connecteur logiciel & Aucune & & Indifférent \\
\hline
Courbe d'apprentissage & Aucune & & Désirable  \\
\hline
\end{tabular}\end{adjustbox}
\end{table}

\subsection{Résultats}

L'exécution du processus automatisé élimine l'alternative Bitcoin, car elle ne permet pas le support de contrats intelligents, ainsi que l'alternative Hyperledger Fabric, car elle ne tolère pas les fautes byzantines. Nous obtenons ainsi deux matrices, l'une contenant les poids et l'autre les alternatives possibles (resp. Ethereum-PoW, Ethereum-PoA, et Corda). Sachant qu'un poids à 0 pour un attribut donné rend celui-ci insignifiant dans le calcul du score de chaque alternative, nous pouvons simplifier ces matrices pour les valeurs suivantes :  

\begin{minipage}[t]{.35\textwidth}
\begin{equation}
     W = \begin{pmatrix} 
     0.25 \\ 0.75 \\ 0.5 \\ 0.5
     \end{pmatrix}
\end{equation}
\end{minipage}
\begin{minipage}[t]{.5\textwidth}
\begin{equation}
    A = \begin{pmatrix} 
    180 & 10 & 1 \\ 0 & 1 & 1 \\ 0.5 & 0.33 & 0.33 \\ 0.4 & 0.4 & 0.8 
    \end{pmatrix}
\end{equation}
\end{minipage}


S'en suit l'exécution de notre algorithme d'aide à la décision, qui propose les résultats suivants (Figure \ref{topsisres}). Notre algorithme de décision considère donc l'alternative Ethereum-PoA comme étant la meilleure. En effet, son score obtenu est le plus proche de 1 (solution idéale positive) des trois alternatives. 

\begin{table}[topsisres]{Résultat de l'exécution du processus de décision.}
\centering
\begin{tabular}{|l|c|}
  \hline
  \textbf{Alternative} & \textbf{Score} \\
  \hline
  Ethereum, PoA & 0.83124114 \\
  Corda, PBFT & 0.71016139 \\
  Ethereum, PoW & 0.28983861 \\
  Hyperledger Fabric, Raft & Disqualifiée \\
  Bitcoin, PoW & Disqualifiée \\
  \hline
\end{tabular}
\end{table}

\subsection{Validation de la solution proposée \label{benchsection}}

Nous avons montré dans la sous-section précédente que la solution la plus adaptée est Ethereum-PoA. Pour confirmer la pertinence de la solution, nous allons expérimenter la robustesse du réseau Ethereum par le biais d'un outil permettant de tester ses performances développé dans ce sens. 

Pour cela, nous avons implémenté un contrat intelligent pour Ethereum qui, lorsque déployé sur la blockchain, permet de faire les opérations définies dans le scénario de chaîne logistique (sauvegarde de données hachées, administration des tiers autorisés à utiliser l'application). 
Nous avons ensuite déployé une blockchain Ethereum-PoA sur Grid'5000, qui est un banc d'essai flexible, de grande taille et configurable à souhait pour le support d'expériences de large échelle.
Les noeuds composant la blockchain possèdent chacun un processeur Intel Xeon Gold 5220 (18 cores), 96 GiB de mémoire vive, deux SSD de 480GB et 960GB respectivement, et une bande passante de 2x25 Gbps. Le nœud chargé de piloter l'expérience par l'envoi de transaction possède les mêmes caractéristiques techniques.
Chacun des nœuds utilise le client d'Ethereum Geth, configuré avec l'algorithme PoA Clique\footnote{https://github.com/ethereum/EIPs/issues/225}, un intervalle de génération de bloc laissé à la valeur par défaut de 5 secondes, et une taille de bloc non limitée.

\begin{figure}[benchresults]{Tests de performance Ethereum-PoA pour le cas d'utilisation supply-chain \newline 
Un client soumet à un réseau Ethereum-PoA contenant $s$ noeuds, $y$ transactions par seconde pendant $n$ secondes. Après $n + \epsilon$ secondes (où $\epsilon$ étant égal à 3 fois la durée inter-blocs), le nombre $z$ de transactions acquittées est retourné. \newline L'expérience est répétée 10 fois.
}
    \centering
    \includegraphics[scale=0.5]{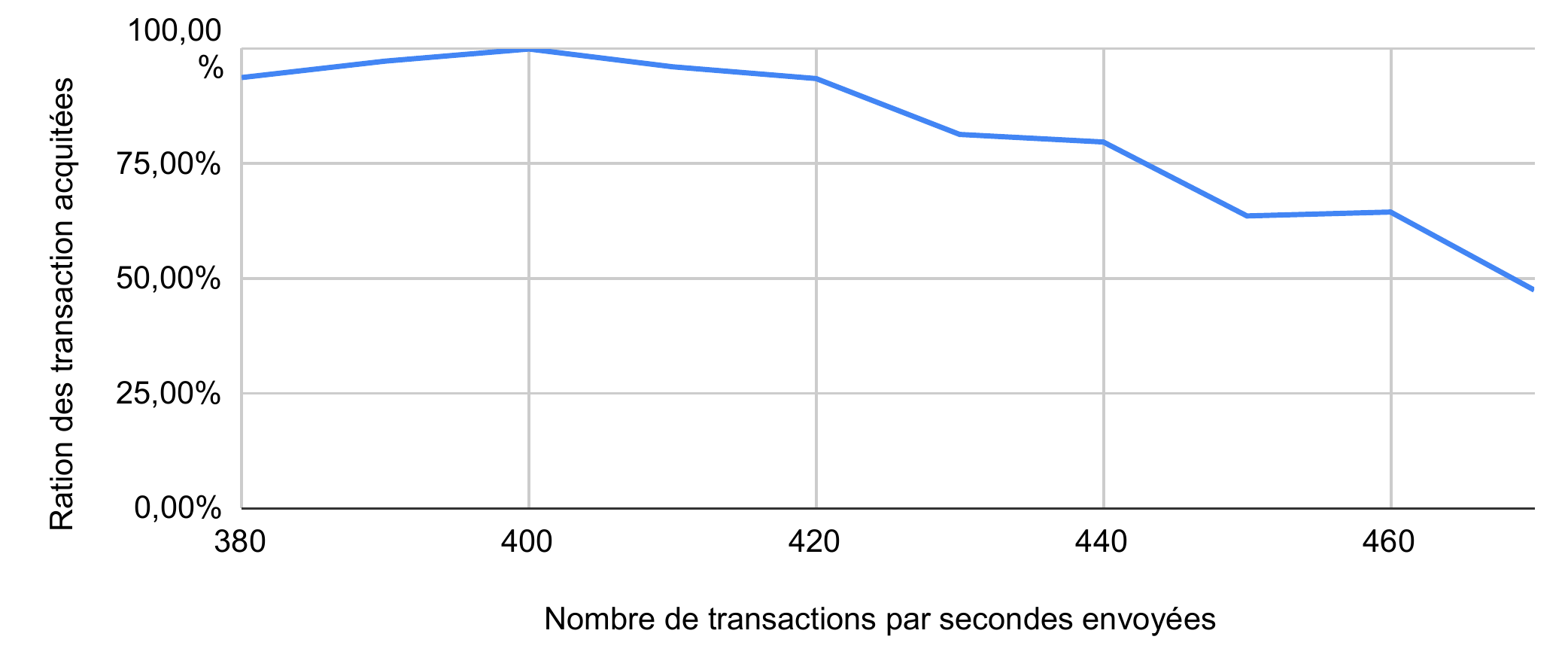}
\end{figure}

Nous constatons que la blockchain arrive bien à supporter une charge de 380 transactions par seconde. Nous en déduisons qu'une telle infrastructure est amplement capable de supporter une charge composée de 60 transactions par jour ainsi que quelques transactions ponctuelles d'administration du consortium. Le choix d'Ethereum-PoA est donc pertinent pour le cas d'utilisation donné.


\section{Travaux connexes \label{relatedwork}}

Nos travaux s'inscrivent dans la lignée d'études réalisées pour faciliter l'adoption de la blockchain par le biais d'une aide à la décision entre différents types de blockchains, ou par la décision entre l'utilisation d'une blockchain ou non dans un contexte donné. 

Dans \cite{Wust2018} les auteurs listent les principales propriétés de la blockchain (Transparence, intégrité, confiance ...) et proposent un modèle de décision sur l'adoption de la blockchain ou non en fonction de la réponse à certaines questions (telles que : "Y a-t-il plusieurs tiers impliqués ?" ou "Sont-ils de confiance ?") liées au cas d'étude donné. 
Ils appliquent ensuite leur modèle à plusieurs cas d'usages d'exemples.
Bien qu'il y ait une étude des paramètres de la blockchain permettant de définir les questions du modèle de décision, le résultat est d'un niveau d'abstraction très élevé (blockchain publique, privée, permissionnée ou pas de blockchain). 
Il ne permet donc pas de prendre une décision précise sur la technologie de blockchain à utiliser ainsi que ces paramètres. 
Dans \cite{Koens2018}, les auteurs effectuent une revue de littérature sur des études relatives aux modèles de décision pour la blockchain afin de construire leur propre modèle. 
Les résultats de celui-ci sont un peu plus précis que le précédent, mais ne donnent toujours pas une recommandation précise. 
Les auteurs de \cite{Labazova2019} ont également réalisé une revue de littérature tout en utilisant une approche DSR (Design Science Research) afin de construire leur modèle. 
Celui-ci comporte plusieurs niveaux de décision et prennent en compte des propriétés blockchain, ce qui permet à un utilisateur de faire un choix avec une précision accrue en sortie par rapport aux études précédentes. 
De plus, les auteurs montrent les dépendances entre certains paramètres (par exemple, la confidentialité et la transparence). Cependant, les paramètres en entrée sont majoritairement spécifiques à la blockchain et conditionnent l'utilisation du modèle par un expert.
Une autre étude intéressante présente une troisième approche d'aide à la décision en proposant un travail complet de détail des fondamentaux blockchain dans la première partie de leur étude, ainsi qu'un modèle de décision introduisant des critères opposés (tels que performance/coûts), mais également une série de questions pour affiner le choix ("Quand utiliser la blockchain ?", "Quoi utiliser ?", "Comment utiliser cette blockchain ?") \cite{Belotti2019}.
Toutes ces études permettent de guider la prise de décision pour un projet blockchain donné, mais ne permettent pas d'aller plus en détail (paramètres blockchain) à cause des limitations des modèles de décision. Le manque d'automatisation et la résolution manuelle des questions ne permettent pas de prendre en compte un grand nombre d'exigences en entrée.

Certaines études ont été réalisées afin de répondre à cette problématique. À titre d'exemple, les auteurs de \cite{Tang2019} proposent d'utiliser une méthode d'aide à la décision multicritère appelée TOPSIS, qui est la même que celle utilisée dans cette étude, afin de déterminer la meilleure solution de blockchain publique disponible à partir d'un ensemble de critères en entrée. L'approche est intéressante dans ce contexte, mais ne permet pas de prendre en compte d'autres blockchains (privées, permissionnées). De plus, les critères techniques blockchain sont regroupés sous les critères "basic technology", "applicability" et "transaction per second", le premier étant quantifié via des experts, les recommandations données en résultat peuvent donc manquer de précision si l'on se place du point de vue de l'entreprise souhaitant démarrer son projet.

Dans \cite{Farshidi2020}, les auteurs ont construit un système de prise de décision pour les technologies blockchain basées sur des travaux précédents pour d'autres technologies. Ils ont réalisé un sondage auprès d'experts pour déterminer les critères de choix les plus pertinents, puis ont rempli une base de connaissance contenant les valeurs de ces attributs choisis pour un ensemble large de blockchains (obtenus avec des livres blancs, études, tests de performance ...)  afin de donner des recommandations via un moteur d'inférence. Leur outil est très performant et permet de donner des recommandations précises, nous voulons aller plus loin en proposant quelque chose plus orienté blockchain (prise en compte de processus métiers et de modèles architecturaux spécifiques) qui soit plus accessible pour des non-experts en blockchain, par le biais d'un modèle qui lie les attributs blockchain et qualité logicielle. Ainsi l'utilisateur peut saisir des exigences plus courantes que celles spécifiques à la technologie blockchain. 

\section{Discussion\label{discussion}}

La prédiction obtenue, qui est d'utiliser Ethereum-PoA, nous semble un choix pertinent pour plusieurs raisons. En effet, toutes les fonctionnalités que nous estimons nécessaires à la bonne implémentation du cas d'étude choisi sont présentes, tout en permettant de garantir un coût optimal de celle-ci (faible difficulté d'apprentissage et économique en énergie). 
Cependant, la méthode demeure sensible aux variations de poids. Si nous avions choisi un poids supérieur concernant le débit de transactions, nous aurions pu avoir un résultat différent en sortie. Des études de sensibilité peuvent permettre d'établir des intervalles, servant à indiquer à quel degré un poids peut varier sans affecter le résultat final. Il existe également des méthodes, comme celle de la détermination de poids par l'entropie, permettant de limiter l'impact des critères ayant une forte entropie en diminuant leur poids \cite{huang2008combining}. Par ailleurs, l'échelle de Likert que nous avons choisi pour l'expression des préférences peut entraîner un biais selon la perception des écarts entre les différentes valeurs proposées par l'utilisateur. Afin de rendre le résultat plus fiable, d'autres systèmes de pondération pourraient être considérés (AHP).

Pour notre seconde expérience mettant en œuvre un test de performance de la blockchain Ethereum-PoA, nous avons trouvé que celle-ci n'était plus capable de traiter 100\% des transactions entrantes à partir de 400 transactions par seconde. Le suivi de l'exécution sur chacun des nœuds montre que cette incapacité apparaît lorsque le CPU des noeuds n'est plus capable de supporter la charge de transactions reçues par le client Geth.
Il est cependant possible de diminuer l'intervalle inter-blocs afin d'augmenter les performances, mais une valeur trop basse pourrait dégrader la qualité du réseau (difficulté à aboutir à un consensus entre nœuds d'autorité) et augmenter l'espace disque nécessaire (chaque bloc comportant au moins un entête de taille non nulle). Par conséquent, nous avons choisi de conserver la valeur par défaut, mais étudier l'impact d'une baisse sur la stabilité pourrait être profitable. Aussi, nous avons constaté lors de notre expérimentation que la courbe représente fidèlement la perte de transactions, mais nous pensons que répéter l'expérimentation pour chaque point de mesure plusieurs fois et allonger le temps de chaque expérience pourrait affiner grandement les résultats. 

\section{Conclusion et travaux futurs \label{conclusion}}

Dans cette étude, nous avons adapté une méthode d'aide à la décision multicritère afin de concevoir un processus de décision automatisé pour blockchain. 
Pour cela, nous avons sélectionné un panel pertinent de blockchains ainsi que de critères relatifs à la qualité d'un système (norme ISO 25010) pour créer une base de connaissance, puis nous avons choisi une liste de termes permettant à un utilisateur de soumettre ses préférences et exigences quant aux critères choisis pour la décision. Enfin, nous avons validé notre processus sur un cas d'étude de gestion de chaîne logistique et montré que notre outil est capable de recommander une blockchain alignée aux besoins de l'utilisateur. L'implémentation est en cours, et sera complétée puis mise à disposition en accès ouvert sur Github dans de futurs travaux.
Cette étude est une première étape pour concevoir un processus de décision automatisé plus étendu, car il pourrait prendre en compte un plus grand nombre d'entrées (topologie d'architecture système, infrastructure, processus métiers...). Cela nous permettrait, à l'aide de ces informations, d'exécuter un test de performance personnalisé (tel que celui présenté dans la sous-section \ref{benchsection}) pour chaque utilisateur avant même d'exécuter l'algorithme de décision, le but étant de fixer de manière extrêmement précise les valeurs des critères variants (débit de transactions, latence ...). 
Une autre piste d'amélioration est l'utilisation d'approches basées sur la logique floue ou les modèles bayésiens qui permettrait de tenir compte de l'aspect subjectif des critères de décision. 

\acknowledgements{
  \footnotesize{
Les expériences présentées dans ce document ont été réalisées à l'aide du banc d'essai Grid'5000, soutenu par un groupe d'intérêt scientifique hébergé par INRIA et comprenant le CNRS, RENATER et plusieurs autres universités et organisations (voir https://www.grid5000.fr).}
}

\bibliography{inforsidBib}

\begin{thebibliography}{}

\bibitem[\protect\citeauthoryear{%
Abeyratne%
\ \BBA{} Monfared%
}{%
Abeyratne%
\ \BBA{} Monfared%
}{%
{\protect\APACyear{2016}}%
}]{%
abeyratne2016blockchain}%
\APACinsertmetastar{%
abeyratne2016blockchain}%
Abeyratne\ S\BPBI A.%
\BCBT{}\ \BBA{} Monfared\ R\BPBI P.%
%
\unskip\
\newblock
\APACrefYearMonthDay{2016}{}{}.
\newblock
\BBOQ{}\APACrefatitle{Blockchain ready manufacturing supply chain using
  distributed ledger}{Blockchain ready manufacturing supply chain using
  distributed ledger}.\BBCQ{}
\newblock
\APACjournalVolNumPages{International Journal of Research in Engineering and
  Technology}{5}{9}{1--10}.
\PrintBackRefs{\CurrentBib}

\bibitem[\protect\citeauthoryear{%
Allen%
\ \BBA{} Seaman%
}{%
Allen%
\ \BBA{} Seaman%
}{%
{\protect\APACyear{2007}}%
}]{%
allen2007likert}%
\APACinsertmetastar{%
allen2007likert}%
Allen\ I\BPBI E.%
\BCBT{}\ \BBA{} Seaman\ C\BPBI A.%
%
\unskip\
\newblock
\APACrefYearMonthDay{2007}{}{}.
\newblock
\BBOQ{}\APACrefatitle{Likert scales and data analyses}{Likert scales and data
  analyses}.\BBCQ{}
\newblock
\APACjournalVolNumPages{Quality progress}{40}{7}{64--65}.
\PrintBackRefs{\CurrentBib}

\bibitem[\protect\citeauthoryear{%
Androulaki%
\ \protect\BOthers{.}}{%
Androulaki%
\ \protect\BOthers{.}}{%
{\protect\APACyear{2018}}%
}]{%
androulaki2018hyperledger}%
\APACinsertmetastar{%
androulaki2018hyperledger}%
Androulaki\ E.%
, Barger\ A.%
, Bortnikov\ V.%
, Cachin\ C.%
, Christidis\ K.%
, De~Caro\ A.%
\BCBL{}\ \BOthersPeriod{.}%
\unskip\
\newblock
\APACrefYearMonthDay{2018}{}{}.
\newblock
\BBOQ{}\APACrefatitle{Hyperledger fabric: a distributed operating system for
  permissioned blockchains}{Hyperledger fabric: a distributed operating system
  for permissioned blockchains}.\BBCQ{}
\newblock
\BIn{} \APACrefbtitle{Proceedings of the Thirteenth EuroSys
  Conference}{Proceedings of the thirteenth eurosys conference}\ \unskip,
  \BPGS\ 1--15.
\PrintBackRefs{\CurrentBib}

\bibitem[\protect\citeauthoryear{%
Belotti%
\ \protect\BOthers{.}}{%
Belotti%
\ \protect\BOthers{.}}{%
{\protect\APACyear{2019}}%
}]{%
Belotti2019}%
\APACinsertmetastar{%
Belotti2019}%
Belotti\ M.%
, Bozic\ N.%
, Pujolle\ G.%
\BCBL{}\ \BBA{} Secci\ S.%
%
\unskip\
\newblock
\APACrefYearMonthDay{2019}{}{}.
\newblock
\BBOQ{}\APACrefatitle{{A Vademecum on Blockchain Technologies: When, Which and
  How}}{{A Vademecum on Blockchain Technologies: When, Which and How}}.\BBCQ{}
\newblock
\APACjournalVolNumPages{IEEE Communications Surveys {\&} Tutorials}{}{}{1--1}.
\PrintBackRefs{\CurrentBib}

\bibitem[\protect\citeauthoryear{%
Brown%
\ \protect\BOthers{.}}{%
Brown%
\ \protect\BOthers{.}}{%
{\protect\APACyear{2016}}%
}]{%
brown2016corda}%
\APACinsertmetastar{%
brown2016corda}%
Brown\ R\BPBI G.%
, Carlyle\ J.%
, Grigg\ I.%
\BCBL{}\ \BBA{} Hearn\ M.%
%
\unskip\
\newblock
\APACrefYearMonthDay{2016}{}{}.
\newblock
\BBOQ{}\APACrefatitle{Corda: an introduction}{Corda: an introduction}.\BBCQ{}
\newblock
\APACjournalVolNumPages{R3 CEV, August}{1}{}{15}.
\PrintBackRefs{\CurrentBib}

\bibitem[\protect\citeauthoryear{%
Ekblaw%
\ \protect\BOthers{.}}{%
Ekblaw%
\ \protect\BOthers{.}}{%
{\protect\APACyear{2016}}%
}]{%
ekblaw2016case}%
\APACinsertmetastar{%
ekblaw2016case}%
Ekblaw\ A.%
, Azaria\ A.%
, Halamka\ J\BPBI D.%
\BCBL{}\ \BBA{} Lippman\ A.%
%
\unskip\
\newblock
\APACrefYearMonthDay{2016}{}{}.
\newblock
\BBOQ{}\APACrefatitle{A Case Study for Blockchain in Healthcare:“MedRec”
  prototype for electronic health records and medical research data}{A case
  study for blockchain in healthcare:“medrec” prototype for electronic
  health records and medical research data}.\BBCQ{}
\newblock
\BIn{} \APACrefbtitle{Proceedings of IEEE open \& big data
  conference}{Proceedings of ieee open \& big data conference}\ \unskip,
  \BVOL~13, \BPG~13.
\PrintBackRefs{\CurrentBib}

\bibitem[\protect\citeauthoryear{%
{Farshidi}%
\ \protect\BOthers{.}}{%
{Farshidi}%
\ \protect\BOthers{.}}{%
{\protect\APACyear{2020}}%
}]{%
Farshidi2020}%
\APACinsertmetastar{%
Farshidi2020}%
{Farshidi}\ S.%
, {Jansen}\ S.%
, {España}\ S.%
\BCBL{}\ \BBA{} {Verkleij}\ J.%
%
\unskip\
\newblock
\APACrefYearMonthDay{2020}{}{}.
\newblock
\BBOQ{}\APACrefatitle{Decision Support for Blockchain Platform Selection: Three
  Industry Case Studies}{Decision support for blockchain platform selection:
  Three industry case studies}.\BBCQ{}
\newblock
\APACjournalVolNumPages{IEEE Transactions on Engineering Management}{}{}{1-20}.
\PrintBackRefs{\CurrentBib}

\bibitem[\protect\citeauthoryear{%
Herbaut%
\ \BBA{} Negru%
}{%
Herbaut%
\ \BBA{} Negru%
}{%
{\protect\APACyear{2017}}%
}]{%
herbaut2017model}%
\APACinsertmetastar{%
herbaut2017model}%
Herbaut\ N.%
\BCBT{}\ \BBA{} Negru\ N.%
%
\unskip\
\newblock
\APACrefYearMonthDay{2017}{}{}.
\newblock
\BBOQ{}\APACrefatitle{A model for collaborative blockchain-based video delivery
  relying on advanced network services chains}{A model for collaborative
  blockchain-based video delivery relying on advanced network services
  chains}.\BBCQ{}
\newblock
\APACjournalVolNumPages{IEEE Communications Magazine}{55}{9}{70--76}.
\PrintBackRefs{\CurrentBib}

\bibitem[\protect\citeauthoryear{%
Huang%
}{%
Huang%
}{%
{\protect\APACyear{2008}}%
}]{%
huang2008combining}%
\APACinsertmetastar{%
huang2008combining}%
Huang\ J.%
%
\unskip\
\newblock
\APACrefYearMonthDay{2008}{}{}.
\newblock
\BBOQ{}\APACrefatitle{Combining entropy weight and TOPSIS method for
  information system selection}{Combining entropy weight and topsis method for
  information system selection}.\BBCQ{}
\newblock
\BIn{} \APACrefbtitle{2008 IEEE Conference on Cybernetics and Intelligent
  Systems}{2008 ieee conference on cybernetics and intelligent systems}\
  \unskip, \BPGS\ 1281--1284.
\PrintBackRefs{\CurrentBib}

\bibitem[\protect\citeauthoryear{%
Hyv{\"a}rinen%
\ \protect\BOthers{.}}{%
Hyv{\"a}rinen%
\ \protect\BOthers{.}}{%
{\protect\APACyear{2017}}%
}]{%
hyvarinen2017blockchain}%
\APACinsertmetastar{%
hyvarinen2017blockchain}%
Hyv{\"a}rinen\ H.%
, Risius\ M.%
\BCBL{}\ \BBA{} Friis\ G.%
%
\unskip\
\newblock
\APACrefYearMonthDay{2017}{}{}.
\newblock
\BBOQ{}\APACrefatitle{A blockchain-based approach towards overcoming financial
  fraud in public sector services}{A blockchain-based approach towards
  overcoming financial fraud in public sector services}.\BBCQ{}
\newblock
\APACjournalVolNumPages{Business \& Information Systems
  Engineering}{59}{6}{441--456}.
\PrintBackRefs{\CurrentBib}

\bibitem[\protect\citeauthoryear{%
Koens%
\ \BBA{} Poll%
}{%
Koens%
\ \BBA{} Poll%
}{%
{\protect\APACyear{2018}}%
}]{%
Koens2018}%
\APACinsertmetastar{%
Koens2018}%
Koens\ T.%
\BCBT{}\ \BBA{} Poll\ E.%
%
\unskip\
\newblock
\APACrefYearMonthDay{2018}{}{}.
\newblock
\BBOQ{}\APACrefatitle{{What Blockchain Alternative Do You Need?}}{{What
  Blockchain Alternative Do You Need?}}\BBCQ{}
\newblock
\BIn{} \unskip, \BPGS\ 113--129.
\newblock
\APACaddressPublisher{}{Springer}.
\PrintBackRefs{\CurrentBib}

\bibitem[\protect\citeauthoryear{%
Kornyshova%
\ \BBA{} Salinesi%
}{%
Kornyshova%
\ \BBA{} Salinesi%
}{%
{\protect\APACyear{2007}}%
}]{%
kornyshova2007mcdm}%
\APACinsertmetastar{%
kornyshova2007mcdm}%
Kornyshova\ E.%
\BCBT{}\ \BBA{} Salinesi\ C.%
%
\unskip\
\newblock
\APACrefYearMonthDay{2007}{}{}.
\newblock
\BBOQ{}\APACrefatitle{MCDM techniques selection approaches: state of the
  art}{Mcdm techniques selection approaches: state of the art}.\BBCQ{}
\newblock
\BIn{} \APACrefbtitle{2007 IEEE Symposium on Computational Intelligence in
  Multi-Criteria Decision-Making}{2007 ieee symposium on computational
  intelligence in multi-criteria decision-making}\ \unskip, \BPGS\ 22--29.
\PrintBackRefs{\CurrentBib}

\bibitem[\protect\citeauthoryear{%
Labazova%
}{%
Labazova%
}{%
{\protect\APACyear{2019}}%
}]{%
Labazova2019}%
\APACinsertmetastar{%
Labazova2019}%
Labazova\ O.%
%
\unskip\
\newblock
\APACrefYearMonthDay{2019}{}{}.
\newblock
\BBOQ{}\APACrefatitle{{Towards a Framework for Evaluation of Blockchain
  Implementations}}{{Towards a Framework for Evaluation of Blockchain
  Implementations}}.\BBCQ{}
\newblock
\APACjournalVolNumPages{ICIS 2019 Proceedings}{}{}{}.
\PrintBackRefs{\CurrentBib}

\bibitem[\protect\citeauthoryear{%
Lai%
\ \protect\BOthers{.}}{%
Lai%
\ \protect\BOthers{.}}{%
{\protect\APACyear{1994}}%
}]{%
lai1994topsis}%
\APACinsertmetastar{%
lai1994topsis}%
Lai\ Y\BHBI J.%
, Liu\ T\BHBI Y.%
\BCBL{}\ \BBA{} Hwang\ C\BHBI L.%
%
\unskip\
\newblock
\APACrefYearMonthDay{1994}{}{}.
\newblock
\BBOQ{}\APACrefatitle{Topsis for MODM}{Topsis for modm}.\BBCQ{}
\newblock
\APACjournalVolNumPages{European journal of operational
  research}{76}{3}{486--500}.
\PrintBackRefs{\CurrentBib}

\bibitem[\protect\citeauthoryear{%
Longo%
\ \protect\BOthers{.}}{%
Longo%
\ \protect\BOthers{.}}{%
{\protect\APACyear{2019}}%
}]{%
longo2019blockchain}%
\APACinsertmetastar{%
longo2019blockchain}%
Longo\ F.%
, Nicoletti\ L.%
, Padovano\ A.%
, d'Atri\ G.%
\BCBL{}\ \BBA{} Forte\ M.%
%
\unskip\
\newblock
\APACrefYearMonthDay{2019}{}{}.
\newblock
\BBOQ{}\APACrefatitle{Blockchain-enabled supply chain: An experimental
  study}{Blockchain-enabled supply chain: An experimental study}.\BBCQ{}
\newblock
\APACjournalVolNumPages{Computers \& Industrial Engineering}{136}{}{57--69}.
\PrintBackRefs{\CurrentBib}

\bibitem[\protect\citeauthoryear{%
Nakamoto%
}{%
Nakamoto%
}{%
{\protect\APACyear{2008}}%
}]{%
nakamoto2008bitcoin}%
\APACinsertmetastar{%
nakamoto2008bitcoin}%
Nakamoto\ S.%
%
\unskip\
\newblock
\APACrefYearMonthDay{2008}{}{}.
\newblock
\APACrefbtitle{Bitcoin: A peer-to-peer electronic cash system.}{Bitcoin: A
  peer-to-peer electronic cash system.}
\PrintBackRefs{\CurrentBib}

\bibitem[\protect\citeauthoryear{%
Szabo%
}{%
Szabo%
}{%
{\protect\APACyear{1997}}%
}]{%
Szabo1997}%
\APACinsertmetastar{%
Szabo1997}%
Szabo\ N.%
%
\unskip\
\newblock
\APACrefYearMonthDay{1997}{9}{}.
\newblock
\BBOQ{}\APACrefatitle{Formalizing and Securing Relationships on Public
  Networks}{Formalizing and securing relationships on public networks}.\BBCQ{}
\newblock
\APACjournalVolNumPages{First Monday}{2}{9}{}.
\PrintBackRefs{\CurrentBib}

\bibitem[\protect\citeauthoryear{%
Takemiya%
\ \BBA{} Vanieiev%
}{%
Takemiya%
\ \BBA{} Vanieiev%
}{%
{\protect\APACyear{2018}}%
}]{%
takemiya2018sora}%
\APACinsertmetastar{%
takemiya2018sora}%
Takemiya\ M.%
\BCBT{}\ \BBA{} Vanieiev\ B.%
%
\unskip\
\newblock
\APACrefYearMonthDay{2018}{}{}.
\newblock
\BBOQ{}\APACrefatitle{Sora identity: secure, digital identity on the
  blockchain}{Sora identity: secure, digital identity on the
  blockchain}.\BBCQ{}
\newblock
\BIn{} \APACrefbtitle{2018 IEEE 42nd Annual Computer Software and Applications
  Conference (COMPSAC)}{2018 ieee 42nd annual computer software and
  applications conference (compsac)}\ \unskip, \BVOL~2, \BPGS\ 582--587.
\PrintBackRefs{\CurrentBib}

\bibitem[\protect\citeauthoryear{%
Tang%
\ \protect\BOthers{.}}{%
Tang%
\ \protect\BOthers{.}}{%
{\protect\APACyear{2019}}%
}]{%
Tang2019}%
\APACinsertmetastar{%
Tang2019}%
Tang\ H.%
, Shi\ Y.%
\BCBL{}\ \BBA{} Dong\ P.%
%
\unskip\
\newblock
\APACrefYearMonthDay{2019}{}{}.
\newblock
\BBOQ{}\APACrefatitle{{Public blockchain evaluation using entropy and
  TOPSIS}}{{Public blockchain evaluation using entropy and TOPSIS}}.\BBCQ{}
\newblock
\APACjournalVolNumPages{Expert Systems with Applications}{117}{}{204--210}.
\PrintBackRefs{\CurrentBib}

\bibitem[\protect\citeauthoryear{%
Tapscott%
}{%
Tapscott%
}{%
{\protect\APACyear{2016}}%
}]{%
dontapscott2016}%
\APACinsertmetastar{%
dontapscott2016}%
Tapscott\ D.%
%
\unskip\
\newblock
\APACrefYear{2016}.
\newblock
\APACrefbtitle{Blockchain Revolution: How the Technology Behind Bitcoin Is
  Changing Money, Business, and the World}{Blockchain revolution: How the
  technology behind bitcoin is changing money, business, and the world}.
\newblock
\APACaddressPublisher{}{Portfolio}.
\PrintBackRefs{\CurrentBib}

\bibitem[\protect\citeauthoryear{%
Wood%
\ \protect\BOthers{.}}{%
Wood%
\ \protect\BOthers{.}}{%
{\protect\APACyear{2014}}%
}]{%
wood2014ethereum}%
\APACinsertmetastar{%
wood2014ethereum}%
Wood\ G.%
\BCBT{}\ \BOthersPeriod{.}%
\unskip\
\newblock
\APACrefYearMonthDay{2014}{}{}.
\newblock
\BBOQ{}\APACrefatitle{Ethereum: A secure decentralised generalised transaction
  ledger}{Ethereum: A secure decentralised generalised transaction
  ledger}.\BBCQ{}
\newblock
\APACjournalVolNumPages{Ethereum project yellow paper}{151}{2014}{1--32}.
\PrintBackRefs{\CurrentBib}

\bibitem[\protect\citeauthoryear{%
Wust%
\ \BBA{} Gervais%
}{%
Wust%
\ \BBA{} Gervais%
}{%
{\protect\APACyear{2018}}%
}]{%
Wust2018}%
\APACinsertmetastar{%
Wust2018}%
Wust\ K.%
\BCBT{}\ \BBA{} Gervais\ A.%
%
\unskip\
\newblock
\APACrefYearMonthDay{2018}{}{}.
\newblock
\BBOQ{}\APACrefatitle{{Do you need a blockchain?}}{{Do you need a
  blockchain?}}\BBCQ{}
\newblock
\APACjournalVolNumPages{Proceedings - 2018 Crypto Valley Conference on
  Blockchain Technology, CVCBT 2018}{}{i}{45--54}.
\PrintBackRefs{\CurrentBib}

\bibitem[\protect\citeauthoryear{%
Xu%
\ \protect\BOthers{.}}{%
Xu%
\ \protect\BOthers{.}}{%
{\protect\APACyear{2018}}%
}]{%
Xu2018}%
\APACinsertmetastar{%
Xu2018}%
Xu\ X.%
, Pautasso\ C.%
, Zhu\ L.%
, Lu\ Q.%
\BCBL{}\ \BBA{} Weber\ I.%
%
\unskip\
\newblock
\APACrefYearMonthDay{2018}{}{}.
\newblock
\BBOQ{}\APACrefatitle{{A pattern collection for blockchain-based
  applications}}{{A pattern collection for blockchain-based
  applications}}.\BBCQ{}
\newblock
\APACjournalVolNumPages{ACM International Conference Proceeding Series}{}{}{}.
\PrintBackRefs{\CurrentBib}

\end{thebibliography}

\end{document}